# Space-Charge Effects

*N. Chauvin*
Commissariat à l'Énergie Atomique et aux Énergies Alternatives, Gif-sur-Yvette, France

**Abstract**
First, this chapter introduces the expressions for the electric and magnetic space-charge internal fields and forces induced by high-intensity beams. Then, the root-mean-square equation with space charge is derived and discussed. In the third section, the one-dimensional Child–Langmuir law, which gives the maximum current density that can be extracted from an ion source, is exposed. Space-charge compensation can occur in the low-energy beam transport lines (located after the ion source). This phenomenon, which counteracts the space-charge defocusing effect, is explained and its main parameters are presented. The fifth section presents an overview of the principal methods to perform beam dynamics numerical simulations. An example of a particles-in-cells code, SolMaxP, which takes into account space-charge compensation, is given. Finally, beam dynamics simulation results obtained with this code in the case of the IFMIF injector are presented.

## 1 Space charge: beam self-generated fields and forces

### 1.1 Space charge

Consider two particles of identical charge $q$. If they are at rest, the Coulomb force exerts a repulsion (see Fig. 1(a)). Now, if they are travelling with a velocity $v = \beta c$, they represent two parallel currents $I = qv$, which attract each other as a result of the effect of their magnetic fields (see Fig. 1(b)).

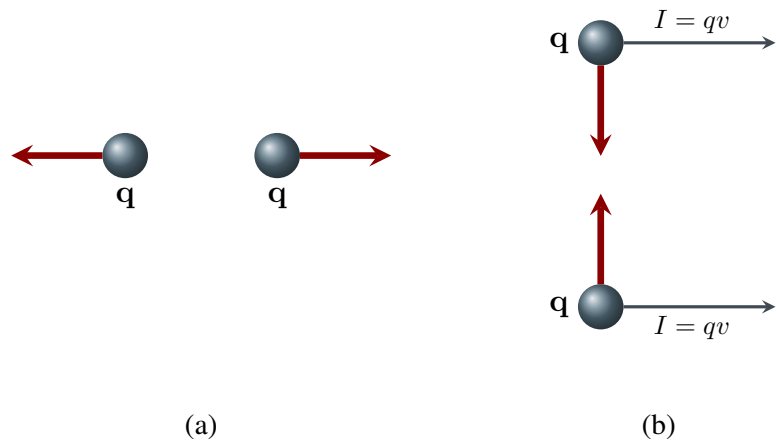

**Fig. 1:** (a) Coulomb force exerts a repulsion between stationary charges. (b) There is attraction between moving charges.

Now, consider a test particle of charge $q$ in an unbunched beam of particles (charge $q_i$) with a circular cross-section (see Fig. 2). The Coulomb repulsion pushes the test particle outwards. The induced force is zero in the beam centre and increases towards the edge of the beam (Fig. 2(a)). The magnetic force is radial and attractive for the test particle in a travelling beam, represented as parallel currents (Fig. 2(b)).

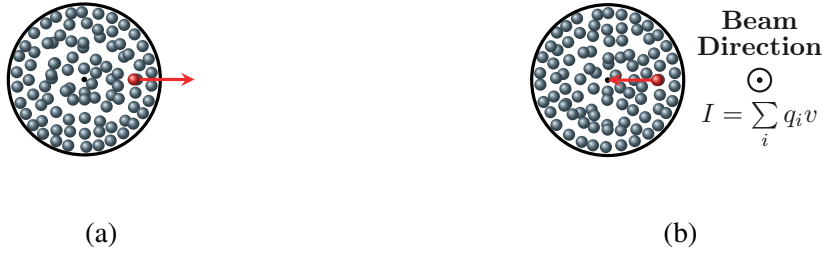

**Fig. 2:** (a) Charges: repulsion. (b) Parallel currents: attraction

## 1.2 Space-charge fields

Consider a continuous beam with a cylindrical symmetry distribution that moves with a constant velocity $v = \beta c$. The beam charge density is

$$\rho(x, y, z) = \rho(r), \tag{1}$$

with $r = \sqrt{(x^2 + y^2)}$. For symmetry reasons, the electric field has only a radial component $E_r$. Using the integral form of Gauss's law over a cylinder centred on the beam axis gives

$$E_r(r) = \frac{1}{\epsilon_0 r} \int_0^r \rho(r') r' \, \mathrm{d}r' \tag{2}$$

The beam current density is

$$\mathbf{J}(x, y, z) = J(r)\mathbf{u}_z, \tag{3}$$

where $\mathbf{u}_z$ is the unitary vector of the beam propagation. If the particles of the beam have the same longitudinal speed, $\mathbf{v}_z = \beta_z c \, \mathbf{u}_z$, we have

$$\mathbf{J}(r) = \rho(r)\beta_z c \, \mathbf{u}_z. \tag{4}$$

For symmetry reasons, the magnetic field has only an azimuthal component $B_\theta$. Using the integral form of Ampere's law over a cylinder centred on the beam axis gives

$$\begin{aligned} B_\theta(r) &= \frac{\mu_0}{r} \int_0^r J(r') r' \, \mathrm{d}r' \\ &= \frac{\mu_0 \beta_z c}{r} \int_0^r \rho(r') r' \, \mathrm{d}r'. \end{aligned} \tag{5}$$

From Eqs. (2) and (5), we obtain (as $c = 1/\sqrt{\epsilon_0 \mu_0}$)

$$B_\theta(r) = \frac{\beta_z}{c} E_r(r). \tag{6}$$

## 1.3 Space-charge forces

The space-charge fields exert a force $\mathbf{F}$ on a test particle at radius $r$ that can be expressed by

$$\mathbf{F} = q(\mathbf{E} + \mathbf{v} \times \mathbf{B}), \tag{7}$$

which with our geometry can be written as

$$F_r = q(E_r + \beta_z c B_\theta). \tag{8}$$

If the particle trajectories obey the paraxial assumption, one can write

$$\beta^2 = \beta_x^2 + \beta_y^2 + \beta_z^2 \approx \beta_z^2. \tag{9}$$

From Eqs. (6) and (8), it follows finally that

$$F_r = qE_r(1 - \beta^2) = \frac{qE_r}{\gamma^2}. \tag{10}$$

The following should be noted about this derivation:

- In Eq. (10), the 1 represents the electric force and the $\beta^2$ the magnetic force.
- The electric force is defocusing for the beam; the magnetic force is focusing.
- The ratio of magnetic to electric forces, $-\beta^2$, is independent of the beam density distribution.
- For relativistic particles, the beam magnetic force almost balances the electric force.
- For non-relativistic particles (like low-energy ion beams), the space magnetic force is negligible: in the ion source extraction region and in the low-energy beam lines, the space charge has a defocusing effect.

### 1.3.1 Uniform beam density

A uniform beam density is expressed by

$$\rho(r) = \begin{cases} \rho_0 & \text{if } r \leq r_0, \\ 0 & \text{if } r > r_0. \end{cases} \tag{11}$$

The charge per unit length is

$$\lambda = \rho_0 \pi r_0^2. \tag{12}$$

The total beam current can be expressed by

$$I = \beta c \int_0^{r_0} 2\pi \rho(r') r' \, dr' \tag{13}$$

So, we obtain

$$\rho_0 = \frac{I}{\beta c \pi r_0^2}. \tag{14}$$

From Eqs. (2) and (14):

$$E_r(r) = \frac{I}{2\pi\epsilon_0 \beta c r_0^2} r \quad \text{if } r \leq r_0, \tag{15a}$$

$$E_r(r) = \frac{I}{2\pi\epsilon_0 \beta c r} \quad \text{if } r > r_0. \tag{15b}$$

The following should be noted:

- The field is linear inside the beam.
- Outside of the beam, it varies according to $1/r$.

Similarly, from Eqs. (5) and (13):

$$B_\theta(r) = \mu_0 \frac{I}{2\pi r_0^2} r \quad \text{if } r \leq r_0, \tag{16a}$$

$$B_\theta(r) = \mu_0 \frac{I}{2\pi r} \quad \text{if } r > r_0. \tag{16b}$$

### 1.3.2 Space-charge forces – Gaussian beam density

A Gaussian beam density is expressed by

$$\rho(r) = \rho_{0\text{g}} \exp\left(\frac{-r^2}{2\sigma_r^2}\right). \tag{17}$$

The charge per unit length is

$$\lambda = 2\rho_{0\text{g}}\pi\sigma_r^2. \tag{18}$$

The space-charge electric field is

$$E_r(r) = \frac{\rho_{0\text{g}}}{\epsilon_0} \frac{\sigma_r^2}{r} \left[1 - \exp\left(\frac{-r^2}{2\sigma_r^2}\right)\right]. \tag{19}$$

The following should be noted:

- The field is nonlinear inside the beam.
- Far from the beam (several $\sigma_r$), it varies according to $1/r$.

## 1.4 Space-charge expansion in a drift region

Consider a particle (charge $q$, mass $m_0$) beam of current $I$, propagating at speed $v = \beta c$ in a drift region, with the following hypotheses:

1. The beam has cylindrical symmetry and a radius $r_0$.
2. The beam is paraxial ($\beta_r \ll \beta_z$).
3. The beam has an emittance equal to 0.
4. The beam density is uniform.

Newton's second law for the transverse motion of the beam particles gives

$$\frac{\mathrm{d}(m_0\gamma\beta_r c)}{\mathrm{d}t} = m_0\gamma\frac{\mathrm{d}^2 r}{\mathrm{d}t^2} = qE_r(r) - q\beta c B_\theta(r). \tag{20}$$

Using Eqs. (15a) for $E_r$ and (16a) for $B_\theta$ in Eq. (20) gives

$$m_0\gamma\frac{\mathrm{d}^2 r}{\mathrm{d}t^2} = \frac{qIr}{2\pi\epsilon_0 r_0^2 \beta c}(1 - \beta^2), \tag{21}$$

with

$$\frac{\mathrm{d}^2 r}{\mathrm{d}t^2} = \beta^2 c^2 \frac{\mathrm{d}^2 r}{\mathrm{d}z^2}. \tag{22}$$

Equation (21) becomes

$$\frac{\mathrm{d}^2 r}{\mathrm{d}z^2} = \frac{qIr}{2\pi\epsilon_0 r_0^2 m_0 c^3 \beta^3 \gamma^3}. \tag{23}$$

We will now introduce several parameters used in the literature on beams with space charge. The characteristic current $I_0$ is defined by

$$I_0 = \frac{4\pi\epsilon_0 m_0 c^3}{q}. \tag{24}$$

The generalized perveance $K$, a dimensionless parameter, is defined by

$$K = \frac{qI}{2\pi\epsilon_0 m_0 c^3 \beta^3 \gamma^3}. \tag{25}$$

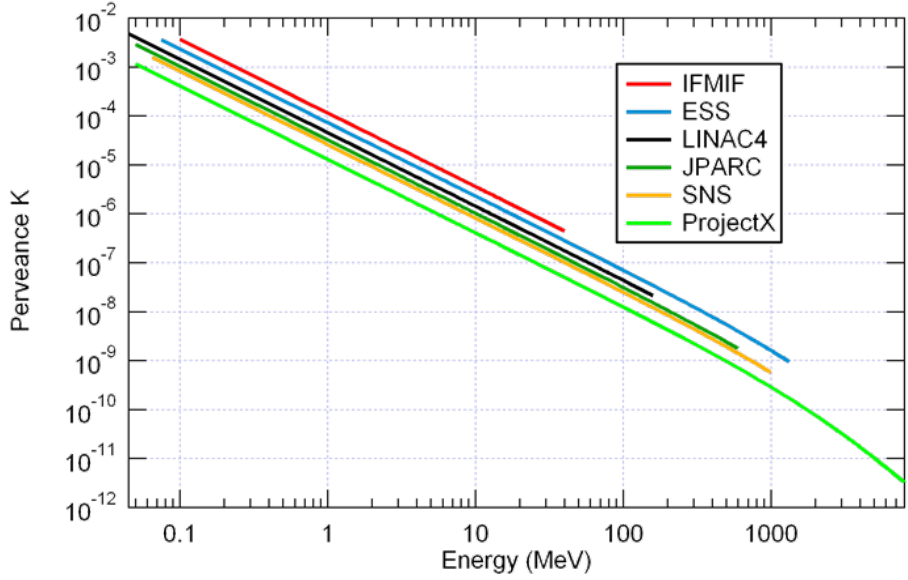

**Fig. 3:** Beam perveance for different hadron accelerators as a function of their beam energy

The perveance refers to the magnitude of space-charge effects in a beam. Figure 3 compares the generalized perveance, as a function of the beam energy, for various high-intensity accelerators.

Then, Eq. (23) for the particle trajectories can be reduced to the form

$$\frac{d^2 r}{dz^2} = \frac{K}{r_0^2} r. \qquad (26)$$

In the case of a laminar beam, the trajectories of all particles are similar and, in particular, the particle at $r = r_0$ will always stay at the beam boundary. Considering $r = r_0 = r_{\text{env}}$, the equation of the beam radius in a drift space can be written as

$$\frac{d^2 r_{\text{env}}}{dz^2} = \frac{K}{r_{\text{env}}}. \qquad (27)$$

## 2 Root-mean-square envelope equation with space charge

### 2.1 Definition of root-mean-square quantities

In the following, $\langle A \rangle$ represents the mean of the quantity $A$ over the beam particle distribution:

$$\text{r.m.s. size} \quad \tilde{x} = \sqrt{\langle x^2 \rangle - \langle x \rangle^2}, \qquad (28)$$

$$\text{r.m.s. divergence} \quad \widetilde{x'} = \sqrt{\langle x'^2 \rangle - \langle x' \rangle^2}, \qquad (29)$$

$$\text{r.m.s. emittance} \quad \widetilde{\epsilon_x} = \sqrt{\tilde{x}^2 \widetilde{x'}^2 - \langle (x - \langle x \rangle)(x' - \langle x' \rangle) \rangle^2}. \qquad (30)$$

The beam Twiss parameters can be expressed with the r.m.s. values as

$$\beta_x = \frac{\tilde{x}^2}{\widetilde{\epsilon_x}}, \qquad (31)$$

$$\gamma_x = \frac{\widetilde{x'}^2}{\widetilde{\epsilon_x}}, \qquad (32)$$

$$\alpha_x = -\frac{\langle (x - \langle x \rangle)(x' - \langle x' \rangle) \rangle}{\widetilde{\epsilon_x}}. \qquad (33)$$

## 2.2 Root-mean-square emittance and nonlinear forces

Consider an ideal particle distribution in phase space $(x, x')$ that lies on a line passing through the origin as represented on Fig. 4. Assume that, for any particles, the divergence $x'$ is related to the position $x$ by the expression

$$x' = Cx^n, \qquad (34)$$

where $n$ is positive and $C$ is a constant. Considering the squared r.m.s. emittance (from Eq. (30))

$$\widetilde{\epsilon}_x^2 = \langle x^2 \rangle \langle x'^2 \rangle - \langle xx' \rangle^2, \qquad (35)$$

and using Eq. (34) we obtain

$$\widetilde{\epsilon}_x^2 = C^2(\langle x^2 \rangle \langle x^{2n} \rangle - \langle x^{n+1} \rangle^2). \qquad (36)$$

When $n = 1$, the line is straight (Fig. 4(a)) and the r.m.s. emittance is $\widetilde{\epsilon}_x^2 = 0$. When $n \neq 1$, the relationship is nonlinear, the line of the phase space is curved (Fig. 4(b)), and the r.m.s. emittance is in general not zero. It is interesting to note that, even if the space phase area occupied by the beam distribution is zero, the r.m.s. emittance may be not zero if the distribution lies on a curved line.

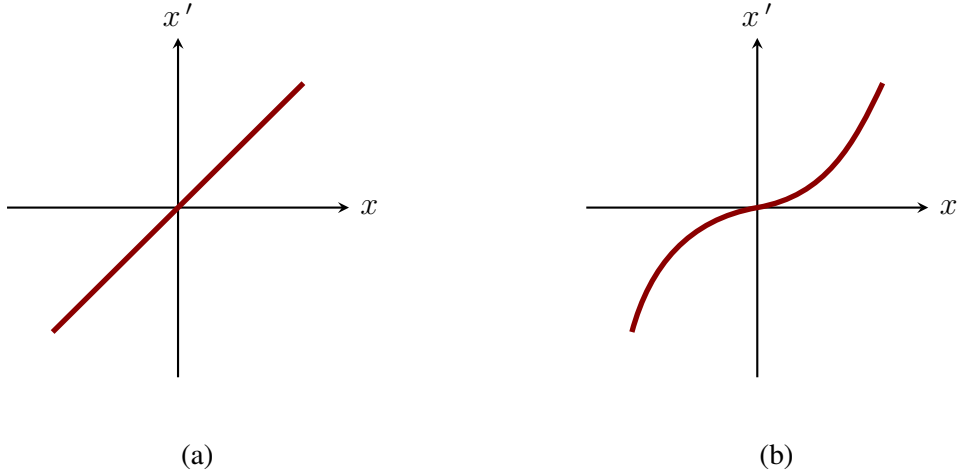

(a)        (b)

**Fig. 4:** Root-mean-square emittance for distributions with zero areas: (a) $\varepsilon = 0$; and (b) $\varepsilon \neq 0$

## 2.3 Root-mean-square envelope equation

Consider a beam moving in the direction $s$, where individual particles satisfy the equation of motion

$$x'' + \kappa(s)x - F_s = 0, \qquad (37)$$

where $\kappa(s)x$ represents a linear external focusing force (for instance quadrupole $\kappa = qB/\gamma ma\beta c$) and $F_s$ is a *space-charge force term* (in general not linear).

To simplify the situation, assume that the beam is centred on the axis with no divergence, so that $\langle x \rangle = 0$ and $\langle x' \rangle = 0$ (i.e. $\tilde{x}^2 = \langle x^2 \rangle \equiv \overline{x^2}$). The equations of motion for the second moments of the distribution can then be written as

$$\frac{\mathrm{d}\overline{x^2}}{\mathrm{d}s} = 2\overline{xx'} \qquad (38)$$

and

$$\begin{aligned}\frac{\mathrm{d}\overline{xx'}}{\mathrm{d}s} &= \overline{x'^2} + \overline{xx''} \\ &= \overline{x'^2} - \kappa(s)\overline{x^2} - \overline{xF_s}.\end{aligned} \qquad (39)$$

Differentiating Eq. (28) twice and using Eq. (38) gives

$$\tilde{x}'' = \frac{\overline{x''} + \overline{x^2}}{\tilde{x}} - \frac{\overline{xx'}}{\tilde{x}^3}. \tag{40}$$

Using Eqs. (39) and (30), we have finally that

$$\tilde{x}'' + \kappa(s)\tilde{x} - \frac{\widetilde{\epsilon_x}^2}{\tilde{x}^3} - \frac{\overline{xF_s}}{\tilde{x}} = 0. \tag{41}$$

Equation (41) is the r.m.s. envelope equation and expresses the motion of the r.m.s. beam size. In three dimensions, one can write three envelope equations that are coupled through the space charge and can depend on the beam dimension in the three directions. These equations can be useful to find an analytic solution for simple problems. Furthermore, they are very important in the space-charge community.

## 2.4 Example: elliptical continuous beam of uniform density

An elliptical uniform beam density is expressed by

$$\rho(r) = \begin{cases} \rho_0 & \text{if } x^2/r_x^2 + y^2/r_y^2 < 1, \\ 0 & \text{otherwise.} \end{cases} \tag{42}$$

As the distribution is uniform, the semi-axes of the ellipse, $r_x$ and $r_y$, are related to the r.m.s. beam sizes: $r_x = 2\tilde{x}$ and $r_y = 2\tilde{y}$. So, the r.m.s. envelope equations are given by

$$\tilde{x}'' + \kappa_x(s)\tilde{x} - \frac{\widetilde{\epsilon_x}^2}{\tilde{x}^3} - \frac{K}{2(\tilde{x}+\tilde{y})} = 0, \tag{43}$$

$$\tilde{y}'' + \kappa_y(s)\tilde{y} - \frac{\widetilde{\epsilon_y}^2}{\tilde{y}^3} - \frac{K}{2(\tilde{x}+\tilde{y})} = 0. \tag{44}$$

The second term of Eqs. (43) and (44) is a focusing term. The third term is the emittance term, which is defocusing (negative sign). The fourth term is the space-charge term, which is also defocusing. It can be seen that the two planes are coupled through the space-charge term of the r.m.s. envelope equation.

## 2.5 Linear space-charge force and equivalent beam

The space-charge force is linear only if the beam density is uniform, which is very unlikely in the case of practical beams. So the space-charge force is fundamentally nonlinear. Nevertheless, it was shown by Sacherer [16] that, for ellipsoidal bunches, where the r.m.s. emittance is either constant or specified in advance, the evolution of the r.m.s. beam projection is nearly independent of the beam density. This means that for calculation of the r.m.s. dynamics, the actual distribution can be replaced by an equivalent uniform beam that has the same intensity and the same r.m.s. second moments.

## 3 The Child–Langmuir law

The Child law states the maximum current density that can be carried by charged-particle flow across a one-dimensional extraction gap. The limit arises from the longitudinal electric fields of the beam space charge. It is a very important result in beam physics of collective effects and also in ion source fields where the extracted beam current is one of the most important parameters of an ion source.

In this section, the Child–Langmuir law for a one-dimensional extraction gap will be derived. It is enough to emphasize the fundamental physical phenomena that limit the beam current extracted from a particle source. However, for a practical extraction system, it is necessary to perform three-dimensional calculations by including form factors that depend on the geometry of the extraction electrodes.

## 3.1 Extraction gap

The extraction gap is the first stage of an accelerator: low-energy charged particles from a source are accelerated to moderate energy (∼10 keV to ∼1 MeV) and formed into a beam. The Child law calculation applies to the one-dimensional gap of Fig. 5. A voltage $-V_0$ is applied across the vacuum gap of width $d$. Charged particles with low kinetic energy enter at the grounded boundary. The particles have rest mass $m_0$ and carry positive charge $q$. Particles leave the right-hand boundary with kinetic energy $qV_0$. We shall assume that the exit electrode is an ideal mesh that defines an equipotential surface while transmitting all particles.

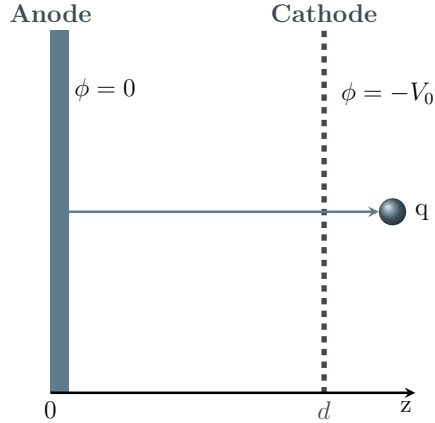

**Fig. 5:** Geometry of an infinite planar extraction gap

## 3.2 One-dimensional Child law for non-relativistic particles

To simplify the calculation, the following assumptions are made:

1. Particle motion is non-relativistic ($qV_0 \ll m_0 c^2$).

2. The source on the left-hand boundary can supply an unlimited flux of particles. Flow restriction would result only from space-charge effects.

3. The transverse dimension of the gap is large compared with $d$.

4. The transverse magnetic force generated by particle current across the gap is negligible compared with the axial electric force. Consequently, the particle trajectories are straight across the gap. This assumption is valid for an ion beam, but can be violated in the case of a high-current relativistic electron gun.

5. Particles flow continuously; the electric field and particle density at all $z$ are constant.

The steady-state condition means that the space-charge density, $\rho(z)$, is constant in time

$$\frac{\partial \rho(z)}{\partial t} = 0. \tag{45}$$

This equation implies that the current density, $J_0$, is the same at all positions in the gap. The charge density can be expressed by

$$\rho(z) = \frac{J_0}{v_z(z)}. \tag{46}$$

According to the above assumptions, the particle velocity in the gap is

$$v_z^2(z) = \frac{2q\phi(z)}{m_0}. \tag{47}$$

Using Eqs. (46) and (47) we obtain

$$\rho(z) = \frac{J_0}{\sqrt{2q\phi(z)/m_0}}. \tag{48}$$

Now, consider the one-dimensional Poisson equation:

$$\nabla^2 \phi = \frac{d^2\phi(z)}{dz^2} = -\frac{\rho(z)}{\epsilon_0}. \tag{49}$$

Substituting Eq. (48) into Eq. (49) gives

$$\frac{d^2\phi(z)}{dz^2} = -\frac{J_0}{\epsilon_0 \sqrt{2q/m_0}} \frac{1}{\phi^{1/2}}. \tag{50}$$

If we introduce the dimensionless variables $\zeta = z/d$ and $\Phi = -\phi/V_0$, Eq. (50) becomes

$$\frac{d^2\Phi(z)}{d\zeta^2} = -\frac{\alpha}{\phi^{1/2}}, \tag{51}$$

with

$$\alpha = \frac{J_0 d^2}{\epsilon_0 V_0 \sqrt{2qV_0/m_0}}. \tag{52}$$

Three boundary conditions are needed to integrate Eq. (51): $\Phi(0) = 0$, $\Phi(1) = 1$ and $d\Phi(0)/d\zeta = 0$. Multiplying both sides of Eq. (51) by $\Phi' = d\Phi/d\zeta$, we can integrate and obtain

$$(\Phi')^2 = 4\alpha\sqrt{\Phi(\zeta)}. \tag{53}$$

A second integration gives

$$\Phi^{3/4} = \tfrac{3}{4}\sqrt{4\alpha}\,\zeta, \tag{54}$$

or, coming back to the initial variables,

$$\phi(z) = -V_0 \left(\frac{z}{d}\right)^{4/3}. \tag{55}$$

In Eq. (54), the condition $\Phi(1) = 1$ implies $\alpha = \tfrac{4}{9}$. Substituting in Eq. (52) gives

$$J_0 = \frac{4}{9}\epsilon_0 \left(\frac{2q}{m_0}\right)^{1/2} \frac{V_0^{3/2}}{d^2}. \tag{56}$$

This is the Child–Langmuir law, which represents the maximum current density that can be achieved in the diode by increasing the ion supply by the anode. This limitation is only due to space charge. The only way to increase the extracted current is to increase the electric field in the gap (i.e. increase the gap voltage or decrease the cathode–anode spacing). In a real ion source, the extracting electric field cannot be increased infinitely, as electrodes break down at fields exceeding a value around 10 MV m$^{-1}$.

For a given gap voltage and geometry, the current density is proportional to the square root of the charge-to-mass ratio of the particles, $\sqrt{q/m_0}$.

Let us compare the electrostatic potential for a space-charge-limited flow given by Eq. (55), with the potential in the same gap without space charge. In that case, the potential between the anode and the cathode of the extraction gap (see Fig. 5) can be calculated from Laplace's equation (as there is no space charge, $\rho = 0$),

$$\nabla^2 \phi = \frac{d^2 \phi(z)}{dz^2} = 0, \quad (57)$$

with the solution

$$\phi(z) = -\frac{V_0}{d} z. \quad (58)$$

By comparing Eq. (55) with Eq. (58), it can be seen that the space charge of the extracted particle lowers the potential (in absolute value) at any given point between the two electrodes of the planar diode.

### 3.3 Child–Langmuir law, orders of magnitudes

The Child–Langmuir current for non-relativistic electrons is

$$J_0 = 2.33 \times 10^{-6} \frac{V_0^{3/2}}{d^2} \quad (\text{A m}^{-2}), \quad (59)$$

where $V_0$ is in volts and $d$ in metres. For ions, Eq. (56) becomes

$$J_0 = 5.44 \times 10^{-8} \sqrt{\frac{Z}{A}} \frac{V_0^{3/2}}{d^2} \quad (\text{A m}^{-2}), \quad (60)$$

where $Z$ is the charge state of the ion and $A$ is its atomic number. For a given extraction voltage and geometry, the possible current density of electrons is around 43 times higher than that of protons.

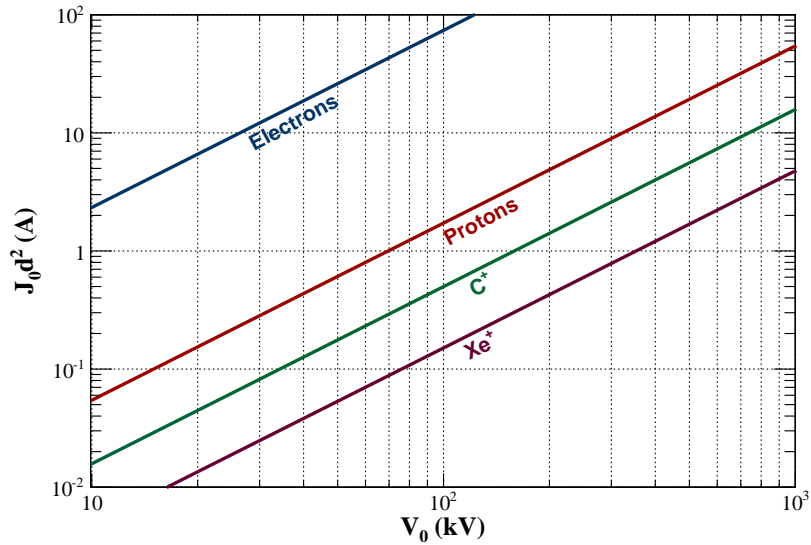

**Fig. 6:** Normalized space-charge-limited current densities for different charged species

Figure 6 shows a plot the variation of $J_0 d^2$ (in amperes) as a function of $V_0$ for electrons and some singly charged ions. The values for electrons are plotted only up to 100 kV, as relativistic corrections should be introduced for higher voltages.

## 4 Space-charge compensation and beam transport in low-energy beam transport lines
### 4.1 Space-charge compensation
#### 4.1.1 Space-charge compensation principle

Space-charge compensation (SCC), or space-charge neutralization, occurs when a beam is propagating through the residual gas of the low-energy beam transport (LEBT) line (or some additional gas) and subsequently induces ionization of the molecules of this gas (Fig. 7). The secondary particles produced by ionization (i.e. electrons or ions), which have an opposite polarity to the particles of the beam, are trapped by the beam potential until a steady state is reached. Thus, the low-energy beam can be considered as a plasma that creates a focusing effect that counteracts the space-charge effect.

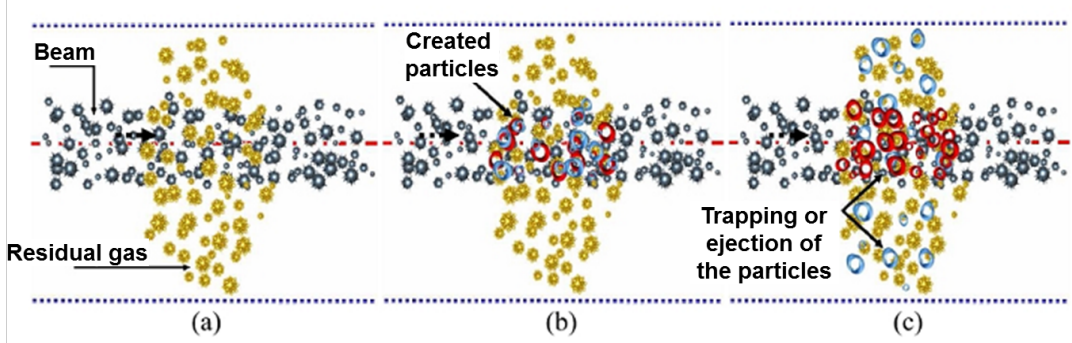

Fig. 7: Space-charge compensation process

As an example, consider a proton beam propagating through a residual gas of $H_2$. It induces the production of $e^-$/$H_2^+$ pairs by ionization according to the following reaction:

$$p + H_2 \to p + e^- + H_2^+ \tag{61}$$

The created electron will be trapped by the beam potential and will contribute to reducing it. The created $H_2^+$ ion will be expelled from the proton beam.

In this section, basic expressions for the degree of space-charge compensation and characteristic time are given to obtain some orders of magnitude for design and experimental considerations. More elaborate descriptions of the SCC evolution can be found in detailed analytical (see [4, 17]) and numerical (see [4, 22]) works.

### 4.1.2 Degree of space-charge compensation

Now, consider a uniform cylindrical beam of radius $r_B$ with an intensity $I_B$ and a longitudinal speed $v = \beta_B c$ propagating into a cylindrical beam pipe of radius $r_P$, which is assumed to be grounded (see Fig. 8).

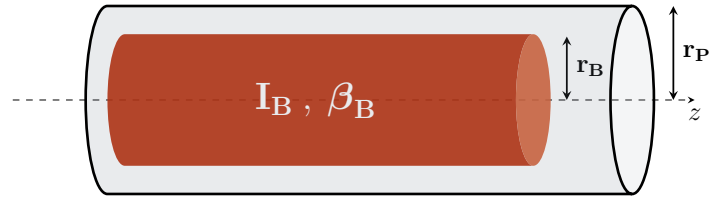

Fig. 8: Uniform cylindrical beam propagating through a beam pipe

If we assume that there is no space-charge compensation, the space-charge electric fields inside and outside the beam are given by Eqs. (15a) and (15b)

$$E_r(r) = \frac{I_B r}{2\pi\epsilon_0 \beta c r_0^2} \quad \text{if } r \leq r_0, \tag{62a}$$

$$E_r(r) = \frac{I_B}{2\pi\epsilon_0 \beta c r} \quad \text{if } r > r_0. \tag{62b}$$

By integrating these equations with the boundary condition, $\phi(r_\mathrm{P}) = 0$, we obtain

$$\phi(r) = \frac{I_\mathrm{B}}{4\pi\epsilon_0 \beta_\mathrm{B} c}\left[1 + 2\ln\left(\frac{r_\mathrm{P}}{r_\mathrm{B}}\right) - \frac{r}{r_\mathrm{B}^2}\right] \quad \text{if } r \leq r_\mathrm{B}, \tag{63a}$$

$$\phi(r) = \frac{I_\mathrm{B}}{2\pi\epsilon_0 \beta_\mathrm{B} c}\ln\frac{r_\mathrm{P}}{r} \quad \text{if } r_\mathrm{B} \leq r \leq r_\mathrm{P}. \tag{63b}$$

The potential on the beam axis (i.e. potential well) created by a uniform beam, without SCC, is given by Eq. (63a) for $r = 0$:

$$\phi_0 = \frac{I_B}{4\pi\varepsilon_0 \beta_B c}\left[1 + 2\ln\left(\frac{r_P}{r_B}\right)\right] \tag{64}$$

During the SCC process, the neutralizing particles created by gas ionization are trapped by this potential well. Equation (64) shows that the potential well (i.e. the space-charge force) increases if the radius of the beam decreases. So, achieving a beam waist in a LEBT could be critical for the quality of the beam.

Now, consider a compensated beam at steady state. If we define by $\phi_\mathrm{c}$ its potential on axis, the SCC degree is given by

$$\eta = 1 - \frac{\phi_\mathrm{c}}{\phi_0}. \tag{65}$$

The beam potential well of the compensated beam can be experimentally measured. The values found for the 75 keV, 130 mA proton beam of the LEDA range from 95% to 99% [9]. Along the LEBT, the SCC degree is not constant, as the neutralizing particle trajectories can be modified by external fields of the focusing element, for example. This phenomenon induces strongly non-uniform space-charge forces that can lead to beam emittance growth

### 4.1.3 Space-charge compensation time

The characteristic SCC transient time, $\tau_\mathrm{scc}$, can be determined by considering the time it takes for a particle of the beam to produce a neutralizing particle on the residual gas. It is given by

$$\tau_\mathrm{scc} = \frac{1}{\sigma_\mathrm{ioniz} n_\mathrm{g} \beta_\mathrm{B} c}, \tag{66}$$

where $\sigma_\mathrm{ioniz}$ is the ionization cross-section of the incoming particles on the residual gas and $n_\mathrm{g}$ is the gas density in the beam line. The space-charge compensation is expected to reach a steady state after two or three $\tau_\mathrm{scc}$ times. As an example, the SCC transient time for a 95 keV proton beam propagating in $H_2$ gas of pressure $5 \times 10^{-5}$ hPa is 15 µs.

### 4.2 Beam transport

Only the transport in the LEBT (Low Energy Beam Transport) will be considered here. Nevertheless, the source extraction system is a critical part, especially for a high-intensity injector, as the beam has to be properly formed under strong space-charge forces to be correctly transported in the LEBT. Thus, it seems mandatory to perform simultaneously both the design and simulations of the ion source extraction and the LEBT.

Once the beam is created and extracted from the ion source, it has to be transported and matched by the LEBT to the first accelerating structure like a radio frequency quadrupole (RFQ). The focus can be achieved with electrostatic or magnetic elements. After the ion source, because of the geometry of the extraction system, the beam usually presents a cylindrical symmetry. In order to preserve this symmetry and to simplify the beam tuning, magnetic solenoid lenses or electrostatic Einzel lenses are more commonly used than quadrupoles.

### 4.2.1 Electrostatic LEBT

In an electrostatic LEBT, the beam is propagating without any space-charge compensation because the neutralizing particles are attracted (or repelled) by the electric field induced by the focusing elements. This kind of beam line is compatible with beam chopping, as there is no transient time for the SCC. Furthermore, the design of electrostatic LEBTs is simplified by the fact that no repelling electrodes for trapping neutralizing particles are needed. So, the beam lines are very compact, which tends to minimize the beam losses by charge exchange. As an example, Fig. 9 shows the SNS ion source with the 12 cm long LEBT equipped with two Einzel lenses [19].

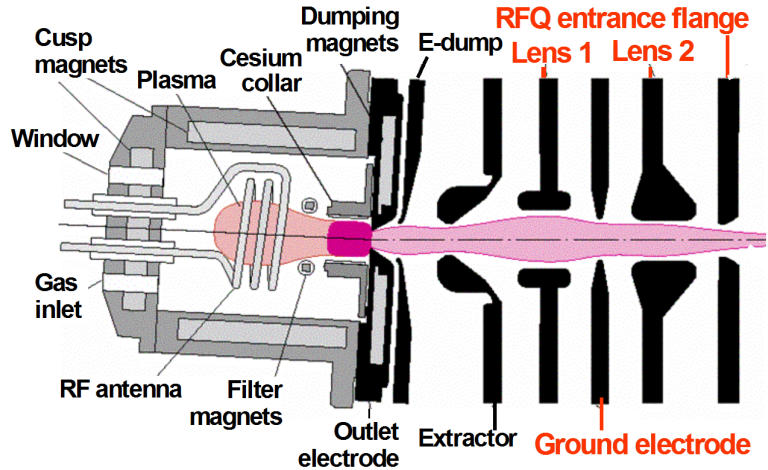

**Fig. 9:** The SNS ion source and electrostatic LEBT

On the down side, electrostatic LEBTs are vulnerable to beam losses that can lead to high-voltage breakdowns and beam trips. Besides, the Einzel lenses intrinsically induce optical aberrations that create a beam halo and emittance growth. To limit this effect, the beam radius should not exceed two-thirds of the lens aperture radius. Finally, the design of the electrostatic LEBTs is intensity-limited. As the beam is not compensated, its divergence and size will increase rapidly with its intensity (especially for currents of several tens of milliamps). So, it seems difficult to operate the LEBT with a higher current than the design current without expecting beam losses or dramatic emittance growth.

### 4.2.2 Magnetostatic LEBT

In this case, the beam is fully compensated by the ionization on the residual gas, as explained in the previous section. The gas in the LEBT comes mainly from the ion source, but it has been shown experimentally that the beam emittance can be improved with a higher pressure in the beam line [11]. Besides, the nature of the injected gas has an influence on this emittance improvement. An emittance reduction of a factor of 2 has been reported by replacing $H_2$ gas by the same partial pressure of Kr [3]. Nevertheless, the gas injection into the beam line has to be done carefully: the higher the pressure, the higher the beam losses by charge exchange. For example, a Kr partial pressure of $4 \times 10^{-5}$ hPa in a 2 m LEBT leads to a $H^+$ (100 keV) loss rate due to electron capture of around 2.4%.

For a positive ion beam, another source of compensation particles should be mentioned, even if it is less significant: secondary electrons are produced when a beam hits the beam pipes.

At the end of the LEBT, the electric field of the RFQ tends to penetrate through the injection hole and have a significant effect on the SCC by attracting the neutralizing particles. Moreover, this region is critical from the space-charge point of view, because a beam waist is performed to match the beam for its injection into the RFQ. So, a polarized electrode is placed as close as possible to the RFQ entrance to repel the neutralizing particles in the LEBT and to minimize the uncompensated zone (see Fig. 10).

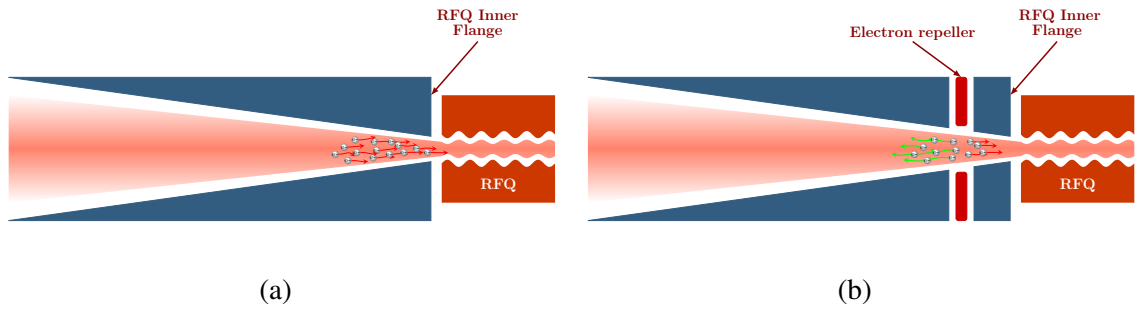

**Fig. 10:** (a) No electron repeller: neutralizing particles (electrons in this case) are attracted by the RFQ electric field. (b) Electrode repeller located before the RFQ: some neutralizing particles are repelled into the LEBT; the uncompensated zone is minimized.

In a magnetic LEBT, the rise time of the pulsed beams is dominated by the SCC transient time (i.e. several tens of microseconds). A fast chopping system has to be inserted to reach a rise time of the order of hundreds of nanoseconds. In the case of H$^-$ ion beams, a phenomenon of overcompensation occurs during the SCC transient time [2]. When the beam is fully compensated, neutralizing particles (in that case H$^+$) are still created, but, as they are significantly slower than the electrons, the SCC degree can be greater than 1 during the time it takes for the excess H$^+$ to be expelled from the beam. During that time, the beam is over-focused and instabilities can be observed.

## 5 Beam dynamics simulation codes with space charge

### 5.1 Numerical codes for ion source extraction systems

Some 2D- or 3D-like codes AXCEL-INP [18], PBGUNS [20] and IBSimu [13] have been successfully used to design sophisticated ion source extraction systems as well as electrostatic LEBTs.

With these codes, one can shape the electrodes, compute the generated electric field and track the particle in the defined domain. Over the past few years, elaborate optimization of the geometry of the extraction system has been performed to increase the extracted beam intensity while minimizing the optical aberration and the beam divergence. As an example, the extraction system of the SILHI source, which has an intermediate and a repelling electrode, forming, together with the plasma and grounded electrodes, a pentode extraction system (see Fig. 11) [7], has been developed using AXCEL-INP.

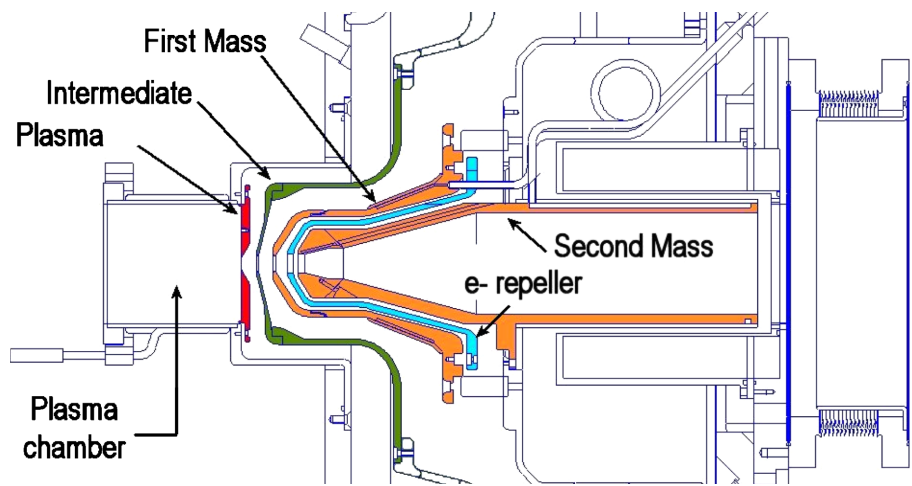

**Fig. 11:** Five-electrode beam extraction system of the SILHI source

## 5.2 Numerical codes for beam dynamics transport

In a classical numerical code, the beam is represented by $N$ macro-particles ($N$ is normally less than the actual particle number in the beam) that can be considered as a statistical sample of the beam with the same dynamics as the real particles. The macro-particles are transported through the accelerator step by step and at each time step $\mathrm{d}t$:

- external forces acting on each macro-particle are calculated,
- space-charge fields and the resulting forces are calculated, and
- tyhe equation of motion is solved for each macro-particle.

The space-charge electric field can be computed by a particle–particle interaction (PPI) method or a particles-in-cells (PIC) method, which are briefly described in the next subsections.

### 5.2.1 Particle–particle interaction method

For each macro-particle $i$ of charge $q$, it is assumed that the applied space-charge electrostatic field, $\vec{E}_i$, is the sum of all the fields induced by all the other macro-particles:

$$\vec{E}_i = \frac{q}{4\pi\varepsilon_0} \sum_{i \neq j} \frac{\vec{r}_j - \vec{r}_i}{\|\vec{r}_j - \vec{r}_i\|^3}. \tag{67}$$

The advantages of this method are that it is easy to code and the electric field is directly computed on the macro-particles. The main drawbacks are that the method is time-consuming for calculation (proportional to the square of the number of macro-particles) and the obtained space-charge field map is not smooth (the lower the macro-particle number, the more granularity), which can lead to non-physical emittance growth.

### 5.2.2 Particles-in-cells method

In this case, the physical simulated space is meshed. The mesh geometry depends on the situation. The meshing can be in one, two or three dimensions, depending on the symmetry of the simulated geometry and the beam.

The average beam density at each node of the mesh is obtained by counting the number of particles that are located close to it (an interpolation can induce smoothing). Once the density function is obtained, the field at each node is computed by solving the Poisson equation. Several techniques can be used to solve this equation at each node of the mesh.

1. A direct method. The field is directly calculated at each node of the mesh. The calculation time is proportional to the square of the mesh number.

2. A fast Fourier transform (FFT) method. The field at one node is given by the convolution product of the density and a Green's function. This can be solved by using the fact that a Fourier transform of a convolution product is equal to the product of the Fourier transforms. If $n$ is the mesh number, the calculation time is proportional to $n \log(n)$. One drawback is that the FFT method does not take into account the boundary conditions of the conductors.

3. A relaxation method [14], which is an iterative method. If $n$ is the mesh number, the calculation time is proportional to $n \log(n)$. It can take into account the particular boundary conditions as conductive items, for instance.

Once the field at each node is known, the field at the macro-particle location is calculated by interpolation from the closest nodes. After the evaluation of the beam density and the field for each particle, the computing time is proportional to the macro-particle number. The algorithm of a PIC code is presented in Fig. 12.

The PIC codes are the most commonly used for space-charge calculations, as they are the fastest and most efficient. A compromise has to be found between the mesh size and the particle number in order reach a sufficient resolution while avoiding some numerical noise that can lead to non-physical effects.

PIC codes that are commonly used for beam dynamics simulations with space charge are: TRACK [1], IMPACT [15] and TraceWin [21].

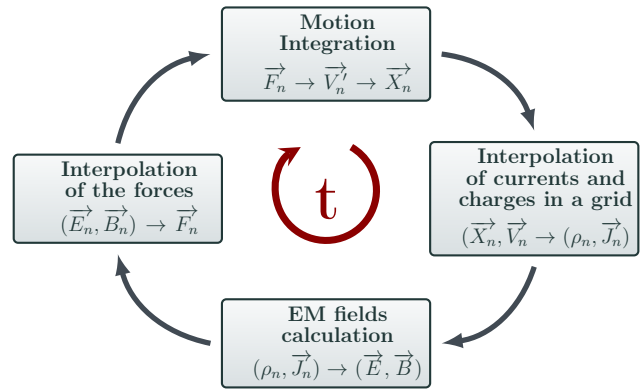

**Fig. 12:** Algorithm of a PIC code dedicated to particle transport with space charge

### 5.3 Beam dynamics numerical codes with space-charge compensation

In order to achieve realistic beam transport simulations of high-intensity ion beams at low energy ($\leq 100$ keV), it is necessary to take into account the space-charge compensation of the beam by ionization of the residual gas. For that, it is necessary to use self-consistent codes, like WARP [12] or SolMaxP [5]. For example, SolMaxP, has recently been developed at CEA-Saclay and is now used to design and simulate high-intensity injectors.

SolMaxP is a PIC code with an additional module (Monte Carlo algorithm) to simulate the beam interactions with gas (ionization, neutralization, scattering) and beam line elements (secondary emission). The dynamics of the main beam is calculated as well as the dynamics of the secondary particles. Figure 13 shows the SolMaxP algorithm.

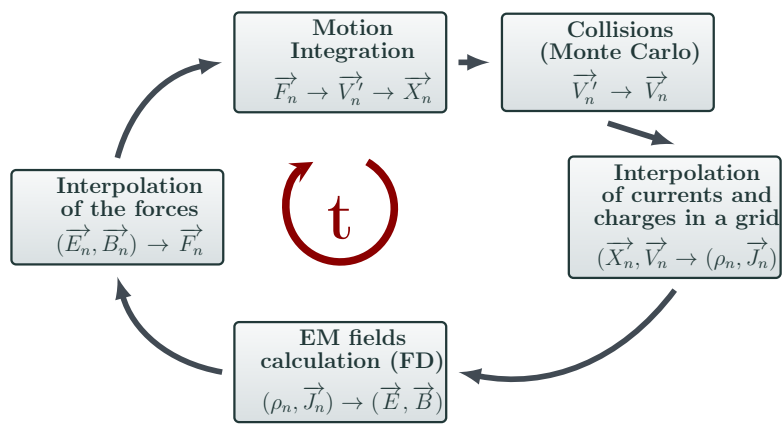

**Fig. 13:** Algorithm of SolMaxP, a self-consistent code for particle transport with space-charge compensation

The SolMaxP code inputs are the particle distribution of the beam, the applied external fields (e.g. focusing elements, electron repeller) and the beam line geometry and gas pressure. The outputs are the particle distributions (ions, electrons, neutral) all along the beam line and the electric field map derived from the potential map created by the space charge along the beam line.

### 5.4 Example of numerical simulations of a high-intensity injector: IFMIF

*5.4.1 Simulation conditions*

First, the modelling of the IFMIF/EVEDA ion source extraction system [8] has been done with AXCEL-INP. The $D^+$, $D_2^+$ and $D_3^+$ particle distributions coming from this model are the input of these simulations.

Then, the simulations presented have been done under the following conditions or hypotheses:

1. $D^+$, $D_2^+$ and $D_3^+$ beams are transported.
2. The electric field map of the source extraction system is included to get relevant boundary conditions.
3. The gas pressure is considered to be homogeneous in the beam line.
4. The gas ionization is produced by ion beam and electron impact.
5. No beam scattering on gas is considered.
6. No secondary electrons are created when the beam hits the beam pipe.

*5.4.2 IFMIF injector parameters and layout*

The IFMIF injector has to deliver a 140 mA, 100 keV CW $D^+$ beam of 0.25 $\pi$ mm mrad emittance. It is composed by a 2.45 GHz electron cyclotron resonance (ECR) source based on the SILHI design and an LEBT with dual solenoid lens focusing system with integrated dipole correcting coils (see Fig. 14) [10].

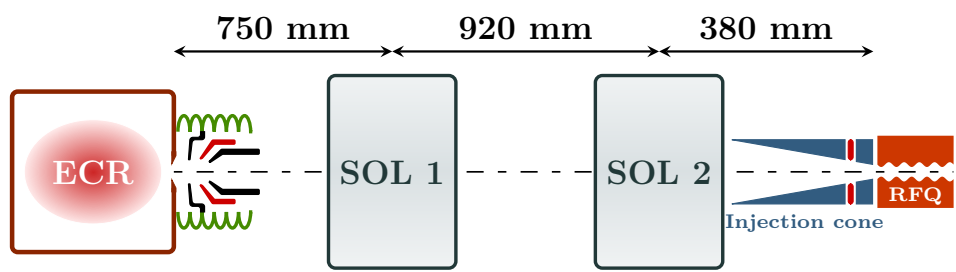

**Fig. 14:** IFMIF source and LEBT layout

*5.4.3 Simulation results*

*5.4.3.1 SCC transient time*

SolMaxP simulations make it possible to determine the SCC transient time. The simulation starts at the time $t = 0$ with no beam in the line but only a fixed gas pressure. Then, the beam starts to propagate in the beam line and its evolution can be followed. The evolution of the $D^+$ beam emittance measured between the two solenoids is represented on Fig. 15. In this plot, the emittance evolution is given for different pressure conditions in the beam line. Assuming that the $D_2$ gas contribution to the total pressure in the beam line is $10^{-5}$ hPa (coming from the ion source), two simulations were done by adding a partial pressure (2 or $4 \times 10^{-5}$ hPa) of either $D_2$ gas or krypton, all the other parameters remaining constant.

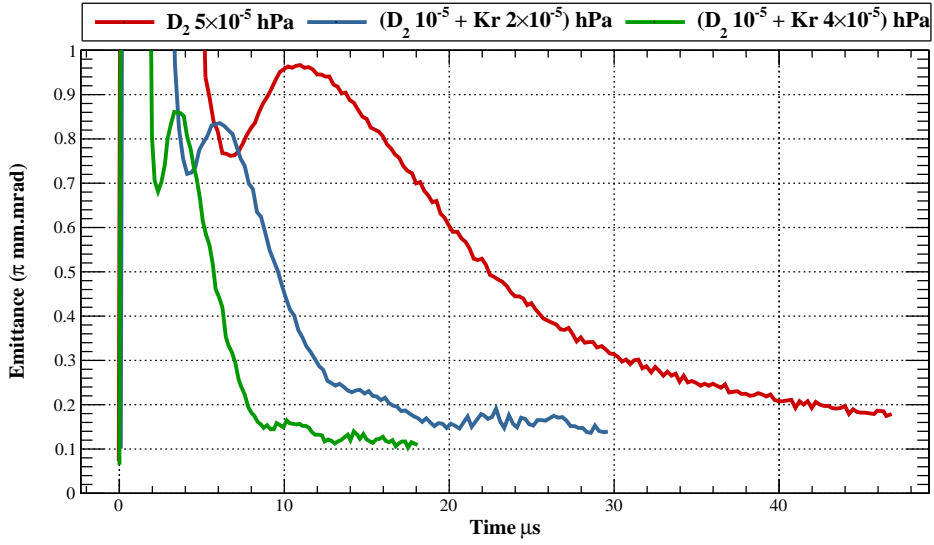

**Fig. 15:** Emittance evolution versus time for different gas pressure in the IFMIF LEBT between the two solenoids

As expected (see Eq. (66)), the SCC transient time decreases when the beam is transported through a gas with a higher ionization cross-section ($\sigma_{\text{ioniz Kr}} > \sigma_{\text{ioniz D}_2}$). Thus, the simulated space-charge compensation transient time is around 20 µs for a krypton pressure of $2 \times 10^{-5}$ hPa.

*5.4.3.2  SCC potential map*

A cut in the $(z0y)$ plane of the space-charge potential in the LEBT, when the space-charge compensation has reached its steady state, is represented on Fig. 16. In this plot, the abscissa $z = 0$ represents the position of the repelling electrode of the source extraction system, while $z = 2.05$ m is the RFQ entrance. The solenoids are respectively located at $z = 0.75$ m and $z = 1.65$ m.

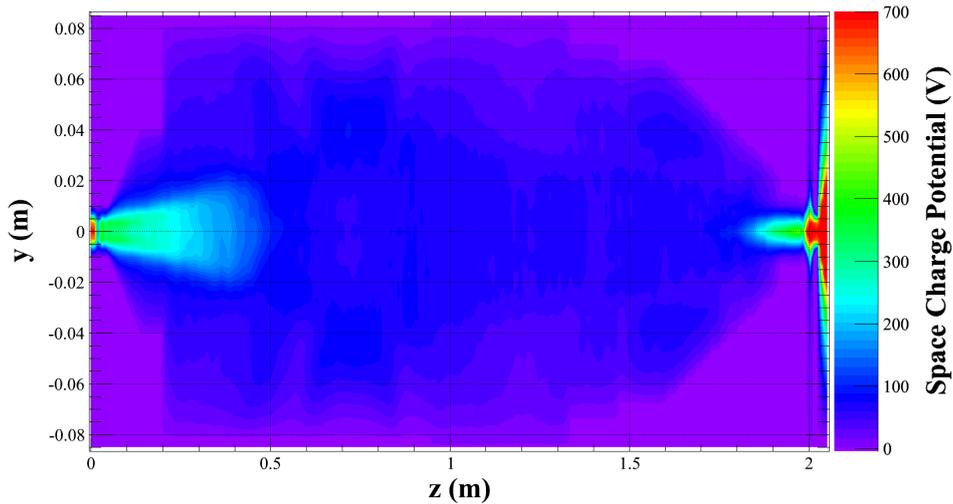

**Fig. 16:** Two-dimensional cut in the $(z0y)$ plane of a space-charge potential map in the IFMIF/EVEDA LEBT

From the space-charge potential map, the SCC degree can be calculated along the IFMIF/EVEDA LEBT. The potential on the uncompensated beam axis $\phi_0$ is calculated with Eq. (65) and the SolMaxP simulations give the potential on the compensated beam axis (see Fig. 16 for $y = 0$). The result is shown in Fig. 17. It can be observed that, in the ion source extraction region and after the repelling electrode at the RFQ injection, the SCC is poor because the electrons are attracted out of the beam. In the central part of the LEBT, where the solenoids and a drift are located, the SCC degree reaches around 95%, which is compatible with the experimentally measured values [9, 11].

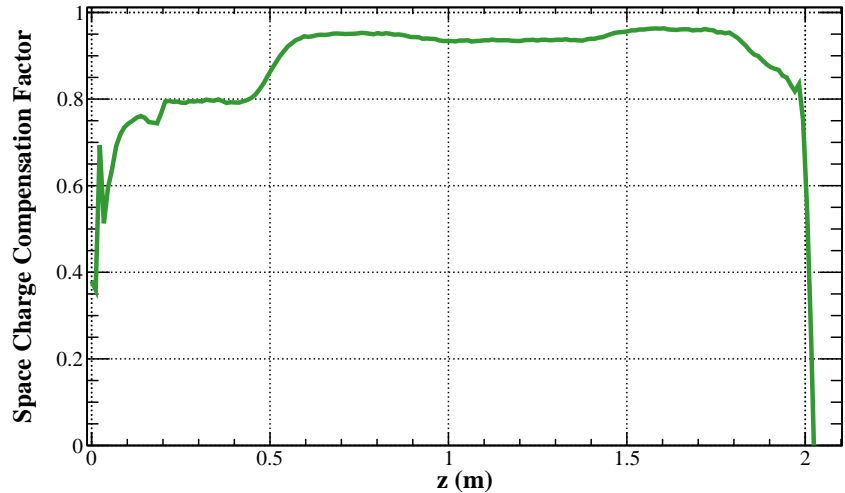

Fig. 17: Space-charge compensation degree along the IFMIF/EVEDA LEBT

*5.4.3.3 IFMIF LEBT beam dynamics*

The extensive calculations that have been performed for the IFMIF injector led to a very compact design of the LEBT and to an optimization of some parameters, like the position of electron repeller electrode in the injection cone [6].

The beam dynamics simulations showed that the IFMIF/EVEDA deuteron beam can be transported and injected into the RFQ with optimized emittance and Twiss parameters. Under these conditions, the RFQ transmission is above 95%. The transport of the $D^+$ beam in the injector and in the first section of the IFMIF RFQ is shown on Fig. 18.

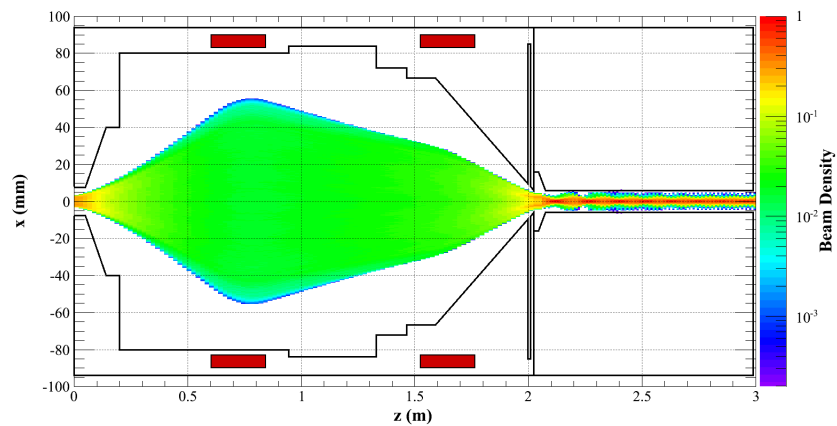

Fig. 18: Beam transport simulation in the IFMIF injector and in the first RFQ cells

The SolMaxP code is in qualitatively good agreement with experimental results, but some quantitative confrontations will be done with experimental results obtained with the beams of the IFMIF injector. Besides, the code will be improved in order to take into account more physical phenomena, like a precise calculation of the gas pressure along the beam line (the pressure appears to be a critical parameter for the SCC) or the beam scattering on the gas.

**Further readings**